\documentclass[
  journal=pasa,
  manuscript=research-paper, 
  year=2025,
  volume=YY,
]{cup-journal}

\usepackage{amsmath}
\usepackage{amssymb,microtype,siunitx,booktabs}
\usepackage{longtable}
\usepackage{tablefootnote}

\title{The radio properties of quasi-periodic X-ray eruption sources}

\author{A. J. Goodwin}
\affiliation{International Centre for Radio Astronomy Research – Curtin University, GPO Box U1987, Perth, WA 6845, Australia}
\email[A.J. Goodwin]{adelle.goodwin@curtin.edu.au}
\author{R. Arcodia}
\affiliation{Kavli Institute for Astrophysics and Space Research, Massachusetts Institute of Technology, Cambridge, MA 02139, USA}
\author{G. Miniutti}
\affiliation{Centro de Astrobiología (CAB), CSIC-INTA, Camino Bajo del Castillo s/n, 28692 Villanueva de la Cañada, Madrid, Spain}
\author{J. C. A. Miller-Jones}
\affiliation{International Centre for Radio Astronomy Research – Curtin University, GPO Box U1987, Perth, WA 6845, Australia}
\author{S. van Velzen}
\affiliation{Leiden Observatory, Leiden University, Postbus 9513, 2300 RA Leiden, The Netherlands}

\received {dd Mmm YYYY}
\revised  {dd Mmm YYYY}
\accepted {dd Mmm YYYY}
\published{22 September 202X}


\begin{document}

\begin{abstract}
    Quasi-periodic X-ray eruptions (QPEs) are a new class of repeating nuclear transient in which repeating X-ray flares are observed coming from the nuclei of generally low mass galaxies. Here we present a comprehensive summary of the radio properties of 12 bona-fide quasi-periodic eruption sources, including a mix of known tidal disruption events (TDEs) and AGN-like hosts. We include a combination of new dedicated radio observations and archival/previously published radio observations to compile a catalogue of radio observations of each source in the sample. We examine the overall radio properties of the sample and compare to the radio properties of known TDEs, given the apparent link between QPEs and TDEs. Overall we find compact, weak radio sources associated with 5/12 of the QPE sources and no signatures of strong AGN activity via a luminous radio jet. 
    We find no radio variability or correlation between radio emission and the X-ray QPE properties, implying that the mechanism that produces the X-ray flares does not generate strong radio-emitting outflows. The compactness of the radio sources and lack of correlation between radio luminosity and SMBH mass is very unusual for AGN, but the radio spectra and luminosities are consistent with outflows produced by a recent TDE (or accretion event), in both the known TDE sources as well as the AGN-like sources in the sample. 
\end{abstract}
 
\section{Introduction} \label{sec:intro}

Quasi-periodic X-ray eruptions (QPEs) are a newly-discovered type of X-ray transient that are characterised by fast and repeating quasi-periodic soft X-ray bursts emitted from the nuclei of low-mass galaxies. They show an X-ray flux increase of more than one order of magnitude over the quiescent plateau with durations of minutes to hours. To date, 12 QPE emitting galaxies have been discovered: GSN 069 \citep{Miniutti2019}, RX J1301.9+2747 \citep{Giustini2020}, eRO-QPE1 \citep{Arcodia2021}, eRO-QPE2 \citep{Arcodia2021}, eRO-QPE3 \citep{Arcodia2024}, eRO-QPE4 \citep{Arcodia2024}, eRO-QPE5 (Arcodia et al in prep), XMMSLJ024916.6--041244 \citep[J0249;]{Chakraborty2021}, ZTF19acnskyy \citep[Ansky;][]{Garcia2025}, and the tidal disruption events AT2019qiz \citep{Nicholl2024}, AT2022upj \citep{Chakraborty2025}, and AT02019vcb \citep{Quintin2023,Bykov2025}. The recurrence time for these sources varies from hours to up a day, and all show a remarkably identical spectral evolution over time during the bursts \citep[e.g.][]{Arcodia2022,Miniutti2023}. 

The physical mechanism that produces QPEs has been proposed to be broadly due to either disc instabilities or interactions between the central SMBH and/or its accretion disk in an extreme mass ratio inspiral (EMRI). In the accretion disk instability scenario, the flares could be due to either instabilties propagating through the disk \citep{Sniegowska2020,Pan2022,Kaur2023} or disk tearing due to disk warping effects \citep{Raj2021}. In the EMRI models, scenarios involving a two-body system consisting of a massive black hole and an orbiting companion with much smaller mass \citep[e.g.][]{King2020,Zhao2022,Wang2022,Linial2023b}, or scenarios in which an orbiting companion undergoing an EMRI passes through a compact pre-existing accretion disk \citep{Arcodia2021,Sukova2021,Linial2023} have been proposed. In the latter scenario, the orbiting body interacts with and shocks the gas in the accretion disk twice on each orbit, which seems to reproduce the observational properties (including the quasi-periodicity of the observed X-ray flares) qualitatively \citep{Lu2023,Linial2023,Franchini2023}. In some of these EMRI models, the accretion disk is remaining from a relatively recent TDE \citep[e.g.][]{Linial2023}. More recently, \citet{Yao2025} showed that repeated star-disk collisions would result in increasing mass loss each collision, suggesting that QPEs are not powered from direct star-disk collisions but instead from stellar debris collisions with the accretion disk. \citet{Mummery2025} showed that star disk collisions with realistic TDE disk models cannot replicate observed QPE flare luminosities, durations, temperatures, and energies, but that a model that invokes stellar debris stream collisions may. 

The connection between QPEs and TDEs was strengthened by the detection of QPEs in the optically-discovered TDEs AT2019qiz \citep{Nicholl2024}, AT2022upj \citep{Chakraborty2025}, and AT2019vcb \citep{Quintin2023,Bykov2025}. Additionally, other QPE sources have shown TDE-like observational properties. Decade-long monitoring of GSN\,069 has shown a long-term X-ray decay including a re-brightening that has been suggested to be due to a partial tidal disruption event \citep{Miniutti2023}, eRO-QPE3 shows a long term X-ray decay \citep{Arcodia2024} and the QPE candidate in \citet{Chakraborty2021} showed a TDE-like multi-wavelength flare prior to the discovery of X-ray variability.


Little is currently known about the radio properties of QPE sources, except for radio detections of GSN 069 presented in \citet{Miniutti2019} and Rx J1302 presented in \citet{Shu2017,Yang2022,Giustini2024}. This is due to the fact that to date not all sources have been observed in the radio, and not all sources were detected if they were observed. TDEs have been found to produce extremely diverse radio emission properties that are well described by outflows ejected by the stellar disruption or accretion process \citep{Alexander2020}. Some radio outflows appear as highly collimated relativistic jets \citep[e.g.][]{Zauderer2011,Pasham2023}, whilst others present slower-moving, likely quasi-spherical outflows \citep[e.g.][]{Alexander2016,vanVelzen2016,Goodwin2022}. The radio emission from TDEs evolves on timescales of months \citep[e.g.][]{Goodwin2024,Goodwin2023,Goodwin2023b, Alexander2016} and in some cases can be delayed by up to years post TDE \citep[e.g.][]{Cendes2022}. However, not all TDEs produce detectable radio emission, with current observations indicating approximately 50$\%$ of events seem to produce detectable radio emission within $\sim5$ years of the TDE \citep{Alexander2020,Cendes2024,Goodwin2025}. 
Given the strengthening link between TDEs and QPEs, it may be expected that QPE sources exhibit similar radio properties to TDEs.


In this work we present a comprehensive summary of the radio properties of 12 bona-fide QPE sources. We include a combination of new dedicated radio observations and archival/previously published radio observations to examine the overall radio properties of these sources and compare to the radio properties of known TDEs. In Section \ref{sec:obs} we present the sample selection and radio observations available for each source, in Section \ref{sec:results} we summarise the radio lightcurves, spectra, and variability statistics for each source, in Section \ref{sec:correlations} we search for any correlations between radio emission and QPE flare and host properties, in Section \ref{sec:discussion} we discuss the implications of these results in the context of interpreting the physical mechanism of QPEs and finally in Section \ref{sec:summary} we summarise the results and provide concluding remarks. 

\section{Sample Selection and Radio Observations}\label{sec:obs}
We searched the literature for confirmed QPE sources. This search resulted in 12 bona-fide QPE sources (where we define ``bona-fide" as a source having two or more QPE flares observed with QPE-like spectral evolution i.e. a harder rise than decay), with flare recurrence times of 2--122\,hr. The 12 sources in our sample and their key properties are listed in Table \ref{tab:source_summary}. 


\begin{table*}[]
\caption{A summary of the properties of the QPE sources studied in this work}
    \centering
    \begin{tabular}{p{1.5cm}llllp{1.5cm}p{1.5cm}p{1.5cm}p{1.5cm}p{2cm}}
\hline
Source & $z$ & $t_{durQPE}$ (ks) & $t_{recurrQPE}$ (ks) & log$_{10}$$E_{QPE}$ (erg) & log$_{10}$$M_{SMBH}$ ($M_{\odot}$)$^*$ & log$_{10}$$M_{gal}$ ($M_{\odot}$) & SFR ($M_{\odot}$/yr) & log$_{10}$$L_{5.5GHz,pk}$ (erg/s) & Ref. \\
\hline
GSN 069 & 0.018 & 4.50$\pm$0.60 & 29.90$\pm$9.50 & 46.11$\pm$0.11 & 6.28$\pm$0.72 & -- & -- &     36.34$\pm$0.07      & \cite{Miniutti2019,Miniutti2023} \\
Rx J1302 & 0.024 & 2.50$\pm$1.10 & 12.90$\pm$10.40 & 45.69$\pm$0.21 & 6.14$\pm$0.88 & -- & -- &     37.441$\pm$0.003     & \cite{Giustini2020,Giustini2024} \\
eRO-QPE1 & 0.051 & 26.60$\pm$5.20 & 77.50$\pm$27.00 & 48.08$\pm$0.12 & 5.90$\pm$0.79 & 9.58$\pm$0.23 & 0.01$\pm$0.00 &     $<$36.99     & \cite{Arcodia2021,Arcodia2022,Chakraborty2024} \\
eRO-QPE2 & 0.018 & 1.70$\pm$0.10 & 8.30$\pm$0.80 & 45.77$\pm$0.04 & 5.43$\pm$0.79 & 9.00$\pm$0.21 & 0.08$\pm$0.07 &     36.87$\pm$0.06      & \cite{Arcodia2021,Arcodia2024b} \\
eRO-QPE3 & 0.024 & 8.40$\pm$0.45 & 72.00$\pm$7.10 & 45.83$\pm$0.13 & 5.53$\pm$0.79 & 9.41$\pm$0.24 & 0.20$\pm$0.14 &     $<$36.59      & \cite{Arcodia2024} \\
eRO-QPE4 & 0.044 & 3.60$\pm$0.40 & 50.80$\pm$10.10 & 46.46$\pm$0.21 & 7.31$\pm$0.75 & 10.20$\pm$0.19 & 2.26$\pm$2.20 &     $<$37.11      & \cite{Arcodia2024} \\
eRO-QPE5 & 0.116 & 51.84$\pm$9.50 & 319.68$\pm$1.73 & 47.53$\pm$0.09 & 7.45$\pm$0.52 & 9.95$\pm$0.18 & -- &     $<$37.86     & Arcodia et al (in prep)\\\
AT2019qiz & 0.015 & 31.90$\pm$1.60 & 175.80$\pm$19.40 & 47.68$\pm$0.04 & 6.27$\pm$0.76 & 10.26$\pm$0.03 & 0.00$\pm$3.50 &     37.33$\pm$0.02      & \cite{Nicholl2024} \\
ZTF19acnskyy & 0.024 & 125.50$\pm$14.10 & 440.90$\pm$85.20 & 47.99$\pm$0.17 & 6.34$\pm$0.66 & -- & -- &     $<$36.81      & \cite{Garcia2025,Sanchez2024} \\
AT2022upj & 0.054 & 59.00$\pm$8.60 & 172.20$\pm$54.20 & 47.77$\pm$0.12 & 6.38$\pm$0.56 & 9.59$\pm$0.11 & -- &     $<$37.96      & \cite{Chakraborty2025,Newsome2024}\\
J0249 & 0.019 & 1.20$\pm$0.05 & 9.50$\pm$1.00 & 45.36$\pm$0.12 & 5.00$\pm$0.50 & 9.10$\pm$0.17 & -- &     $<$37.02      & \cite{Chakraborty2021,Wevers2019} \\
AT2019vcb & 0.088 & 54$\pm$18 & 144$\pm$108 & 48.81$\pm$0.15 & 6.81$\pm$0.13 & 9.49$\pm$0.06 & -- &     37.5$\pm$0.2 & \cite{Quintin2023,Bykov2025} \\
\hline
    \end{tabular}
    $^*$ SMBH mass measurements are obtained from various methods, including scaling with stellar mass, velocity dispersion, and SED fitting.
    $z$ is the source redshift, $t_{durQPE}$ is the average duration of the observed QPEs, $t_{recurrQPE}$ is the average recurrence time of the QPEs, $E_{QPE}$ is the average energy radiated per QPE, $M_{SMBH}$ is the SMBH mass, $M_{gal}$ is the galaxy stellar mass, SFR is the estimated star formation rate of the galaxy, and $L_{5.5GHz,pk}$ is the peak observed 5.5\,GHz radio luminosity. 
    \label{tab:source_summary}
\end{table*}

Of the 12 sources, 8 have radio observations previously published in the literature, we obtained dedicated follow-up radio observations of 3 (including additional radio observations of GSN\,069 to constrain any long-term variability), and we searched archival radio survey data for the remaining 2. A summary of all of the radio observations available for each of the sources is given in the Appendix in Table \ref{tab:radio_obs}. 

\subsection{New radio observations}

\subsubsection{ATCA}
We obtained dedicated radio observations with the Australia Compact Telescope Array (ATCA) of eRO-QPE2 and GSN 069. We observed the coordinates of eRO-QPE2 on three occasions at central frequencies of 2.1, 5.5, and 9\,GHz and the coordinates of GSN 069 on one occasion at a central frequency of 2.1, 5.5, and 9\,GHz. In all observations we used the full 2.048\,GHz of bandwidth split into 2048 spectral channels. All radio observations were reduced using standard procedures in the Common Astronomy Software Application \citep[CASA;][]{CASA2022}, including flux calibration with PKS 1934--638 and phase calibration with PKS 0244--470 (eRO-QPE2) and PKS 0104--408 (GSN069). Images of the target field were created using the CASA task \texttt{tclean} and source flux densities were extracted using the CASA task \texttt{imfit} by fitting a Gaussian the size of the synthesised beam.  Reported flux densities include the statistical error from \texttt{imfit} and a systematic uncertainty estimated to be 5$\%$, added in quadrature. 

GSN\,069 was detected at 2.1, 5.5, and 9\,GHz and showed a point source coincident with nucleus of the host galaxy (Figure \ref{fig:radcontours}). 
eRO-QPE2 was detected at 5.5 and 9\,GHz in the initial observation, however due to the array being in the compact H214 configuration it was confused with a nearby source (see \ref{sec:eRO-QPE2}). We therefore report the measured flux densities for this epoch in Table \ref{tab:radio_obs} as an upper limit on the flux density of the target. In the second ATCA observation with the array in the extended 6\,km configuration, eRO-QPE2 was marginally detected at 5.5 and 9\,GHz and not detected at 2.1\,GHz and did not suffer from confusion with the nearby radio sources. The radio source is consistent with a point source localised to the nucleus of the host galaxy (Figure \ref{fig:radcontours}). In the third ATCA observation, eRO-QPE2 was undetected but the 3$\sigma$ upper limit measured is consistent with the flux density measured in the 2nd epoch. We conclude that there is a weak, compact radio source associated with eRO-QPE2 but we are unable to constrain any variability of the radio source associated with these observations. 

\subsubsection{VLA}
AT2019vcb was observed by the Karl G. Jansky Very Large Array (VLA) in 2020 as part of a dedicated TDE radio follow-up program (program ID 20A-392). The source was observed four times between April 2020 and July 2020 at 2--15\,GHz (S-, C-, X-, and Ku-band). In all observations, 3C286 was used for flux and bandpass calibration and J1221+2813 (S-, C-, X-band) and J1310+3220 (Ku-band) were used for phase calibration. We reduced the data in CASA using standard procedures including the VLA pipeline. Images of the target field were created using the CASA task \texttt{tclean} and when a point source was detected at the location of the target, we used the CASA task \texttt{imfit} to fit a Gaussian the size of the synthesised beam in order to extract the flux density. Reported flux densities include the statistical error from \texttt{imfit} and a systematic uncertainty estimated to be 5$\%$, added in quadrature. A faint, point-like radio source was detected in all observations except at 10 and 15\,GHz in July 2020. The radio flux densities and 3$\sigma$ upper limits are reported in Table \ref{tab:radio_obs}. 

\subsubsection{Archival observations}
We searched the VLA Sky Survey \citep[VLASS;][]{Lacy2020} for observations at the coordinates of the two sources in our sample without dedicated radio observations available: AT2022upj and J0249. There were 2-3 epochs of VLASS observations available for each source, taken between 2017--2023. No radio source was detected coincident with the coordinates of either source in any of the VLASS observations. The 3$\sigma$ VLASS upper limits are reported in Table \ref{tab:radio_obs}. 

We additionally searched the Rapid ASKAP Continuum Survey \citep[RACS][]{McConnell2020} for observations at the coordinates of each of the sources in the sample. All sources were undetected in publicly available RACs observations (with an average image rms of $300-400$\,$\mu$\,Jy) except  RX J1301.9+2747, for which a point source was detected in a 0.88\,GHz RACs observation taken on 2020 December 17. The flux density of this observation is reported in Table \ref{tab:radio_obs}.

\section{Results}\label{sec:results}
5 out of 12 of the QPE sources are radio-detected. The radio-detected sources are faint point-like sources and span a range of luminosities ($\nu L_{\nu}\sim5\times10^{36}-5\times10^{38}$\,erg/s).  

\begin{figure*}
    \centering
    \includegraphics[width=1.1\columnwidth]{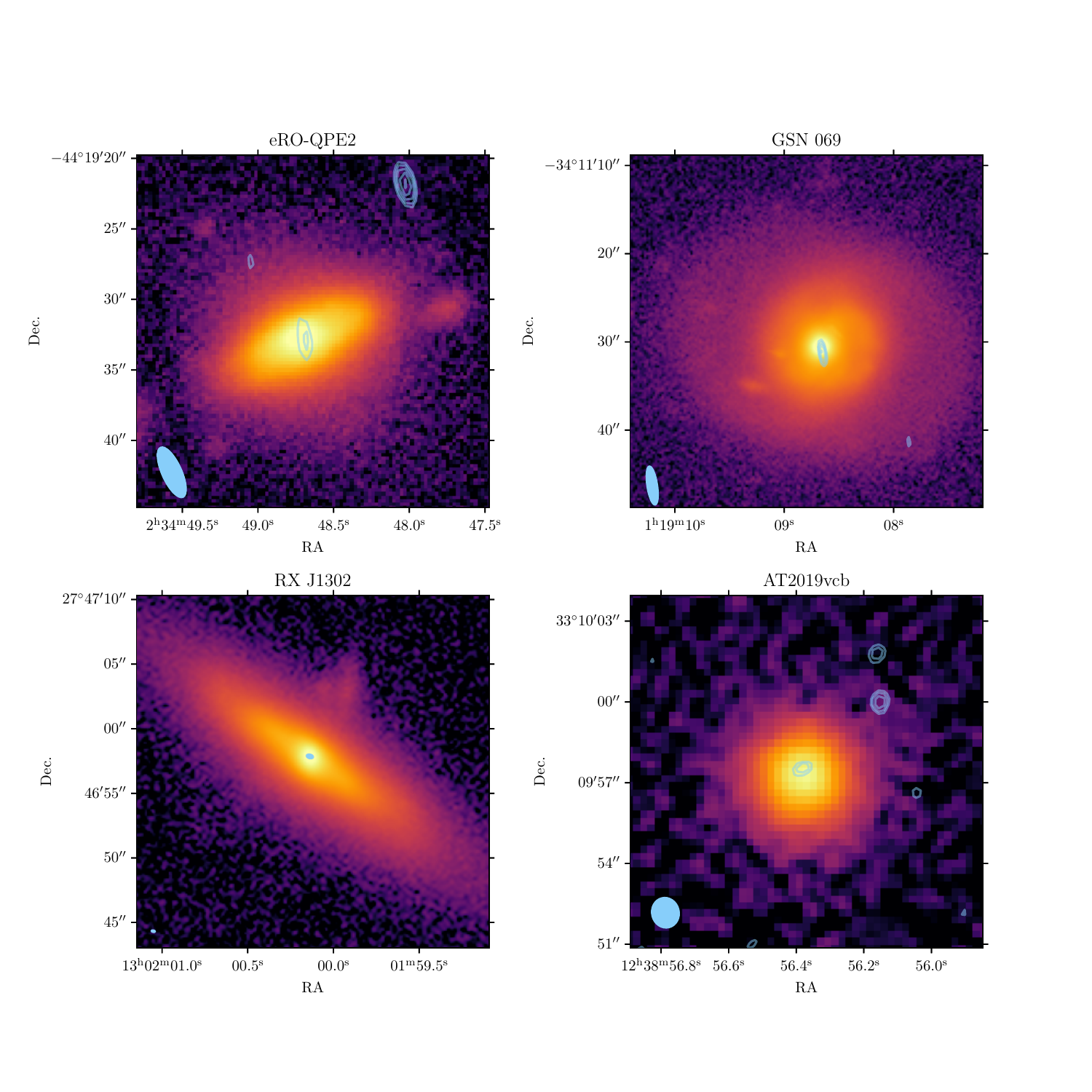}
    \caption{The DESI Legacy Survey DR8-DR10 optical images \citep{Dey2019} of the host galaxies of eRO-QPE2, GSN 069, RX J1302, and AT2019vcb. The ATCA 5.5\,GHz (eRO-QPE2 and GSN 069) and VLA 6\,GHz (RX J1302 and AT2019vcb) radio contours are overlaid in blue, with the radio beam size indicated in the bottom left corner. The radio sources are all compact and localised to the nuclei of the host galaxies.}
    \label{fig:radcontours}
\end{figure*}

The radio contours of eRO-QPE2, GSN069, RXJ1302, and AT2019vcb are plotted over the Legacy Survey DR8--10 \citep{Dey2019} optical images of the host galaxies in Figure \ref{fig:radcontours}. A publicly available image of the AT2019qiz radio observations was not available. It is clear that the radio sources are associated with the nucleus of each of the galaxies, and are not consistent with diffuse star-formation emission in the disk of the galaxies. Although, we cannot rule out faint radio emission from diffuse star formation emission in any of the observations due to the sensitivity of the radio interferometers to faint diffuse emission on these length scales. 

\subsection{Radio lightcurves}
\begin{figure}
    \centering
    \includegraphics[width=\columnwidth]{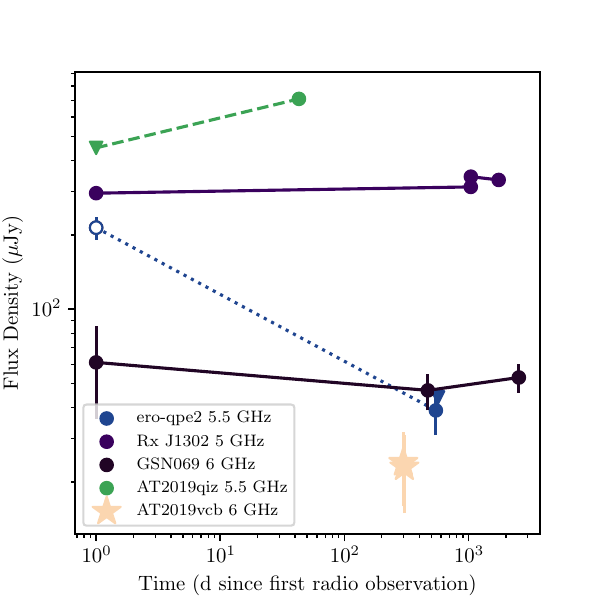}
    \caption{The 5--6\,GHz radio lightcurves of the 5 radio-detected QPE sources in our sample. Inverted triangles indicate 3$\sigma$ upper limits and the open circle indicates the measured flux density affected by confusion with nearby sources. AT2019qiz and RxJ1302 show statistically significant radio variability, while GSN\,069 is constant over the 5\,yr baseline observed. AT2019vcb showed statistically significant variability at 10\,GHz, but the two 6\,GHz observations are only 6\,d apart and do not show significant variability. Our observations of eRO-QPE2 are inconclusive regarding its radio variability (see \ref{sec:eRO-QPE2}).}
    \label{fig:lcs}
\end{figure}

The 5-6\,GHz radio lightcurves of the five radio-detected QPE sources are plotted in Figure \ref{fig:lcs}. AT2019qiz shows rising radio emission over $\approx100$\,d and eRO-QPE2 and AT2019vcb show fading radio emission over a few hundred days, although the first observation of eRO-QPE2 is affected by a confusing source nearby. GSN 069 and RxJ1302 show approximately stable radio sources over $\sim1000$\,d. Previous studies have detected low-amplitude radio variability in GSN 069 \citep{Miniutti2019} and Rx J1302 \citep{Shu2017,Yang2022,Giustini2024}, although no correlation with the X-ray QPE flares was observed. 

\subsection{Radio spectra}

\begin{figure*}
    \centering
    \includegraphics[width=\linewidth]{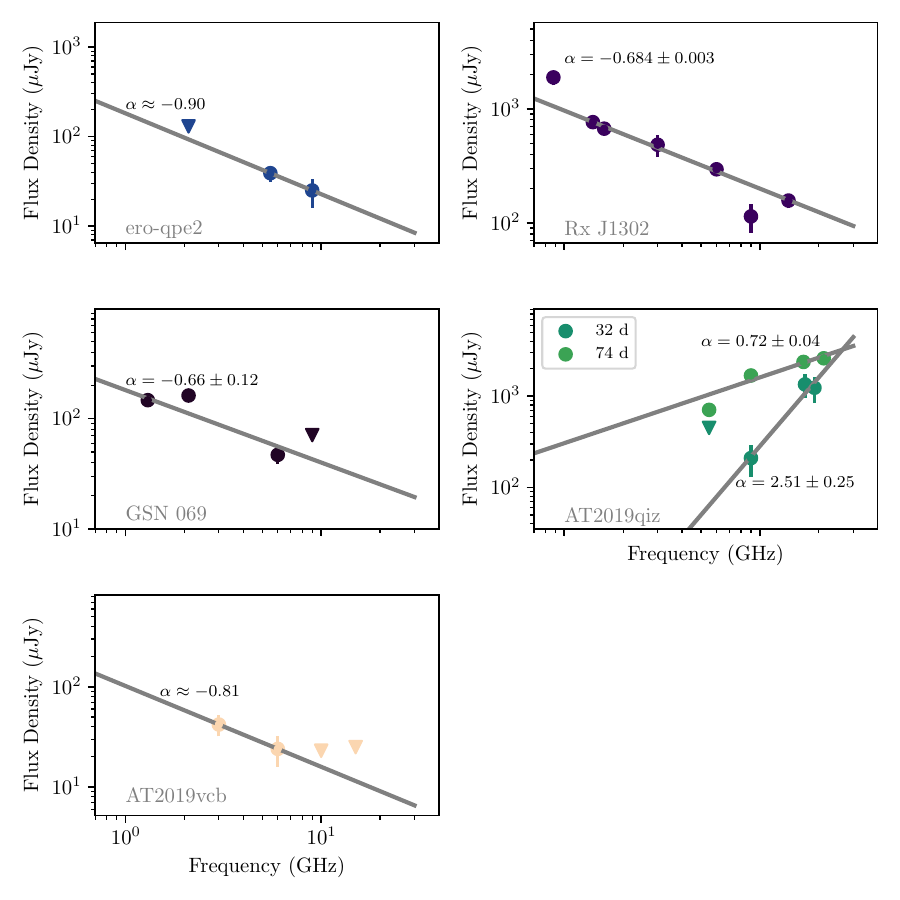}
    \caption{Radio spectra of the four radio-detected QPE sources. In each plot the solid line shows the simple power-law model used to constrain the spectral index, $\alpha$, for each source, where $F_{\nu}\propto\nu^{\alpha}$. Inverted triangles indicate 3$\sigma$ upper limits. For AT2019qiz two spectra are plotted and labelled by days since optical discovery of the TDE. We plot only the ATCA observation taken in December 2023 for eRO-QPE2. }
    \label{fig:specs}
\end{figure*}

We plot the 1--20\,GHz radio spectra of the five radio-detected sources in Figure \ref{fig:specs}. We fit each of the spectra with a simple power-law model

\begin{equation}
    F_{\nu} = A\nu^{\alpha}
\end{equation}
where $A$ is a normalisation constant and $\alpha$ the spectral index. The measured spectral indices for each of the four sources are shown in Figure \ref{fig:specs}. 

AT2019qiz shows an inverted radio spectrum typical of a young TDE that is evolving quickly as the radio spectral peak shifts down in frequency due to the expanding synchrotron-emitting region. The remaining 4 sources show optically thin spectra with spectral indices in the range $-0.7$ to $-0.9$, typical of host galaxy emission from star formation \citep[e.g.][]{Murphy2011}, low luminosity AGN \citep[e.g.][]{Ho2001}, or radio-emitting TDEs \citep[e.g.][]{Goodwin2025}.

\subsection{Radio Variability Statistics}

In order to assess the radio variability properties of the radio-detected sources in our sample, we calculate a variability statistic for the sources in which multiple observations are available at the same observing frequency. We calculate a variability statistic, $V$, such that

\begin{equation}
    V = \frac{(S_{\rm{max}} - \sigma_{S\rm{max}}) - (S_{\rm{min}} + \sigma_{S\rm{min}})}{(S_{\rm{max}} - \sigma_{S\rm{max}}) +(S_{\rm{min}} + \sigma_{S\rm{min}})},
\end{equation}
where $S_{\rm{max}}$ and $S_{\rm{min}}$ are the maximum and minimum observed flux density, and $\sigma_{S\rm{max}}$ and $\sigma_{S\rm{min}}$ their associated uncertainties. Any value of $V>0$ indicates statistically significant variability. The calculated values of $V$ for each of the detected radio sources are given in Table \ref{tab:variability}.  

\begin{table}[]
    \centering
        \caption{The variability statistic, $V$, of the radio emission from each of the radio-detected sources in the sample}
    \begin{tabular}{llll}
     Source Name & Obs. Freq. (GHz) & $V$ & Variable?\\
    \hline
    \hline
        GSN 069 & 6 & -0.21 & No \\
        \hline
        RX J1301.9+2747 & 5 & 0.06 & Yes \\
        & 9 & 0.23 & Yes \\
        \hline
        eRO-QPE2 & 5.5 & $<$0.61 & Unsure$^*$ \\
        \hline
        AT2019qiz & 5.5 & 0.70 & Yes \\
        \hline
        AT2019vcb & 10 & $>$0.33 & Yes \\         
        \hline
    \end{tabular}\\
    $^*$The variability of this source observed may be due to resolution differences in the two observations and a nearby confusing source, see \ref{sec:eRO-QPE2}.
    \label{tab:variability}
\end{table}

GSN\,069 does not show statistically significant variability over the 5\,year baseline probed, whereas RxJ1302, AT2019qiz, and AT2019vcb do show statistically significant variability. Our observations of eRO-QPE2 are inconclusive regarding its radio variability due to a confusing source in the first observation (see \ref{sec:eRO-QPE2}). 

\section{Search for correlations between radio emission and flare/host properties}\label{sec:correlations}

Given the radio sources associated with the 5 radio-detected QPE sources are compact and associated with the nuclei of the host galaxies (Figure \ref{fig:radcontours}), hinting that the radio emission could be due to the QPE mechanism or the mechanism that produced the accretion disk, here we search for any correlation between the measured radio luminosity of the QPE sources and their QPE flare and host properties. 

\begin{figure}
    \centering
    \includegraphics[width=\columnwidth]{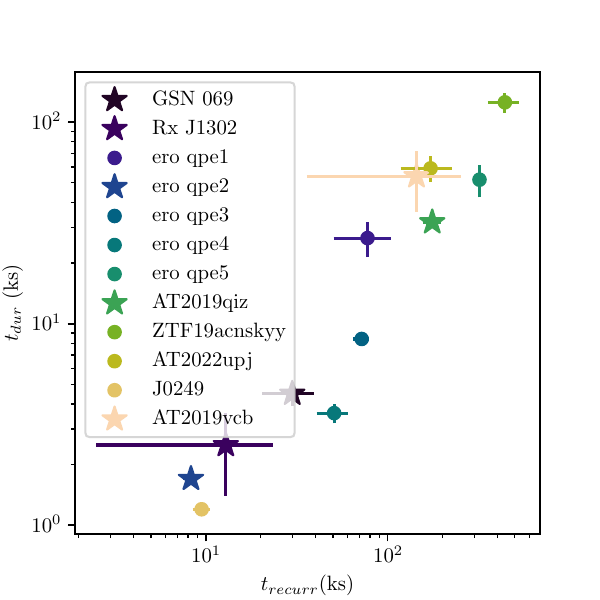}
    \caption{The QPE flare recurrence time ($t_{recurr}$) and flare duration ($t_{dur}$) for each of the 12 QPEs in our sample. Stars indicated QPE sources associated with a compact radio source while circles indicate radio-undetected QPE sources. We find no correlation between the presence of radio emission and the duration or recurrence time of the QPEs.}
    \label{fig:tdur_recurr}
\end{figure}

A strong positive correlation between observed QPE X-ray flare duration ($t_{dur}$) and the average recurrence time ($t_{recurr}$) has previously been established \citep[e.g.][]{Arcodia2024,Mummery2025}. In Figure \ref{fig:tdur_recurr} we plot each of the 12 QPE sources in our sample with the radio-detected sources indicated. In this figure it is evident that the radio-detected QPE sources do not occupy the same region of $t_{dur}$/$t_{recurr}$ parameter space, implying that the radio luminosity is not directly linked to the QPE flare duration or recurrence time. 

\begin{figure*}
    \centering
    \includegraphics[width=\columnwidth]{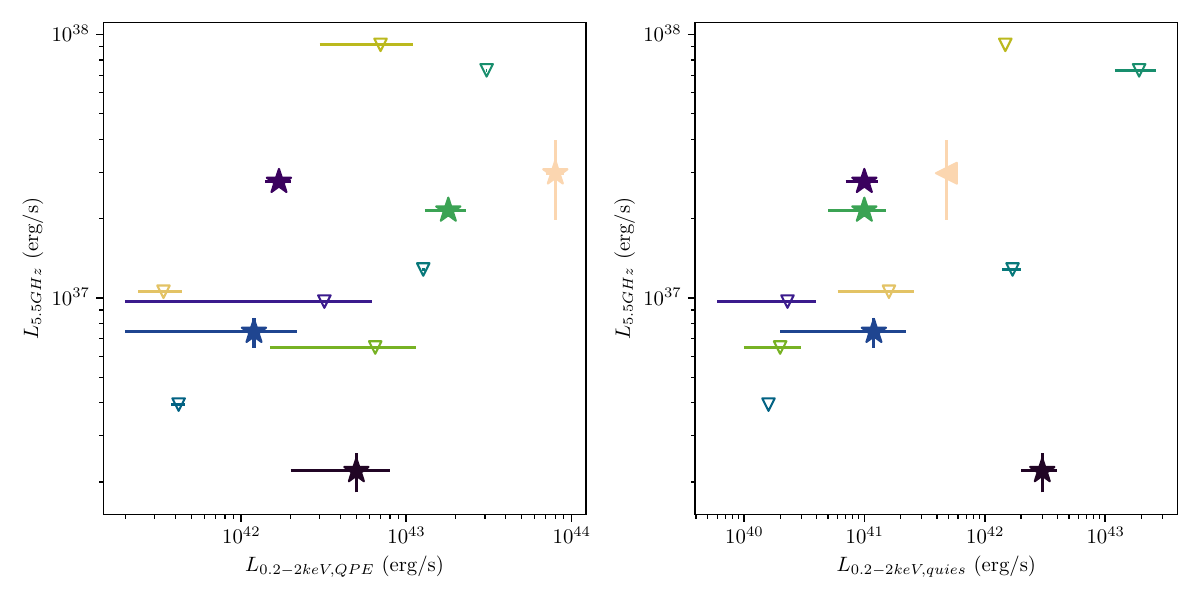}
    \caption{The QPE peak 0.2-2\,keV X-ray luminosity (left panel) and quiescent 0.2-2\,keV X-ray luminosity (right panel) plotted against 5.5\,GHz radio luminosity for the 12 QPE sources in the sample. We find no statistically significant correlation between the radio and X-ray luminosities in the sample.}
    \label{fig:X-rayradL}
\end{figure*}

In Figure \ref{fig:X-rayradL} we plot the observed peak 5.5\,GHz radio luminosities against the observed 0.2-2\,keV QPE and quiescent X-ray luminosities for each of the sources. No correlation between the radio luminosity and either of the X-ray luminosities is visually apparent, and a Pearson correlation test of the radio-detected sources returns a correlation coefficient of 0.58 and p-value of 0.30 (for QPE X-ray luminosity) and correlation coefficient of -0.65 and p-value of 0.23 (for quiescent X-ray luminosity). We therefore do not find a statistically significant correlation between the QPE or quiescent X-ray luminosities and the radio luminosity of the QPE sources. 

\begin{figure*}
    \centering
    \includegraphics[width=\columnwidth]{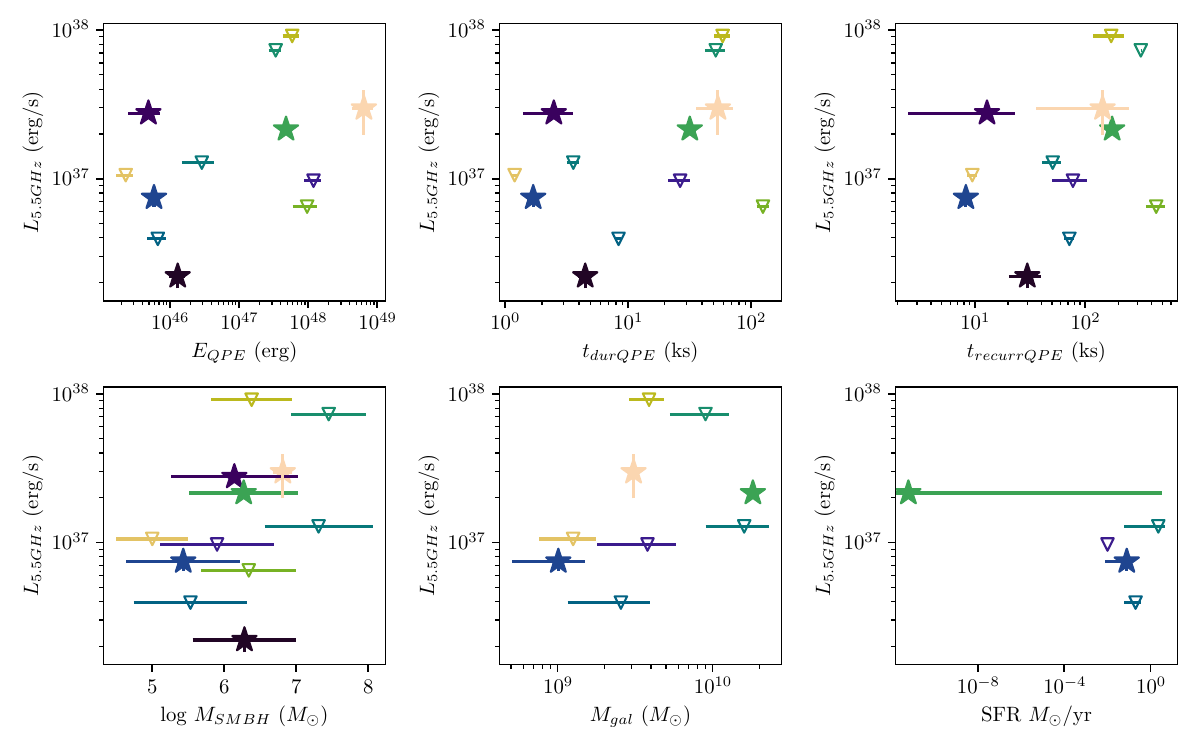}
    \caption{\textit{Top row:} The QPE properties: total energy radiated, duration, and recurrence time plotted against the observed 5.5\,GHz radio luminosity for the 12 QPEs in the sample. We find no correlation between radio luminosity and any of the QPE properties examined. \textit{Bottom row:} The QPE host galaxy properties SMBH mass, galaxy mass, and star formation rate plotted against the observed 5.5\,GHz radio luminosity for the 12 QPEs in the sample. Again we find no correlation between the radio luminosity and the host galaxy properties in our sample. Note that only 5 of the QPE sources had SFRs available in the literature.}
    \label{fig:sample_corrs}
\end{figure*}

We further search for any correlation between QPE flare properties and radio emission by plotting the observed 5.5\,GHz radio luminosity against QPE average energy radiated, duration, and recurrence time in Figure \ref{fig:sample_corrs}.  We find no apparent correlation between radio luminosity and any of the QPE flare properties examined. Performing a Pearson correlation test for the radio-detected sources, we obtain coefficients of 0.5-0.62 with p-values of 0.27--0.38, indicating no statistically significant linear correlation between the radio luminosities and QPE flare properties. 

Next we search for any correlation between the QPE host galaxy properties and the observed radio emission. In Figure \ref{fig:sample_corrs} we plot the observed 5.5\,GHz radio luminosity against QPE host SMBH mass, galaxy mass, and star formation rate (where these properties were available in the literature for each source). We find no apparent correlation between radio luminosity and any of the QPE host properties examined. Performing a Pearson correlation test for the radio-detected sources, we obtain coefficients of 0.25-0.56 with p-values of 0.32--0.68, indicating no statistically significant linear correlation between the radio luminosities and QPE host galaxy properties. 

\section{Discussion}\label{sec:discussion}

Our findings reveal that over half (7/12) of the known QPE sources are not associated with radio sources. Of the 5/12 QPE sources associated with radio sources, the radio emission is faint for radio emission associated with the nuclei of galaxies ($\nu L_{\nu}\sim5\times10^{36}-5\times10^{38}$\,erg/s) and localised to the nucleus of the galaxy (Figure \ref{fig:radcontours}). 

\subsection{The nature of the radio sources}
Given the detected radio sources associated with 5 of the QPE sources are compact and low-luminosity, they may be linked to either the QPE mechanism in the form of a persistent radio-emitting jet or outflow or synchrotron emission from expanding material ejected in each flare, or a compact jet or outflow due to low-level AGN or previous TDE activity. Radio emission from star formation in the host galaxies is an unlikely explanation for the radio emission given the compactness of the radio sources. This conclusion is strengthened by the lack of correlation between radio luminosity and galaxy star formation rate in Figure \ref{fig:sample_corrs}. 

The radio source associated with Rx\,J1302 has been studied the most extensively to date due to it having the highest flux density of the sources in the sample. \citet{Yang2022} showed in extensive VLA observations taken between 2015--2019 that there is significant variability of the radio source on timescales as short as days, implying an emission region size $<10^{-3}$\, pc if the variability is intrinsic to the source. VLBA observations at 1.6\,GHz revealed compact radio emission unresolved on scales $<0.7$\,pc \citep{Yang2022}. \citet{Giustini2024} provided detailed radio monitoring observations of the source taken between 2020--2022 and found significant variability on timescales as short as hours, however in simultaneous XMM-Newton and VLA observations of 5 QPE flares there was no radio variability correlated with the QPEs. \citet{Yang2022} and \citet{Giustini2024} deduce that the stochastic radio variability is consistent with variability induced by interstellar scintillation if the source is $<0.008$\,pc. Such a small source size rules out star formation or pc-scale jets or outflows typically seen in AGN, leaving the possibility of a persistent compact jet, episodic jet ejections, or a compact outflow.   
Simultaneous MeerKAT and Chandra observations of GSN\,069 reported by \citet{Miniutti2019} also showed no radio variability associated with the one QPE flare observed simultaneously, with correlated variability excluded down to the few percent level. In this work, we additionally find no statistically significant radio variability over a 5\,yr baseline, despite the decade-long X-ray decay that has been observed in this source \citep[e.g.][]{Miniutti2023}. 
In Section \ref{sec:correlations} we found no apparent correlation between the observed radio luminosities and the QPE flare properties (Figure \ref{fig:sample_corrs}). 
The lack of correlation between the QPE X-ray variability and radio variability rules out a common mechanism for the radio emission and QPE flares. 

The alternative is that the radio emission is produced by black hole activity, which could take the form of low-level AGN jet activity or an old outflow from a TDE. Whilst AGN commonly have radio spectral indices in the range of those of our radio-detected sample \citep{Ho2001}, AGN are very rarely compact on linear scales $<1$\,kpc \citep[e.g.][]{Ho2001,Anderson2005}. 
In our observations of eRO-QPE2 and GSN\,069, based on the resolution we constrain the sources to be $<2.3"$ and $<5.6"$ respectively, corresponding to linear sizes of $<0.9$\,kpc and $<2.2$\,kpc at 9\,GHz. VLBI observations of RxJ1302 constrained the source to be $<0.7$\,pc \citep{Yang2022}. 
Encountering 3/3 compact persistent radio-detected sources in our sample (excluding the TDEs AT2019qiz and AT2019vcb which show transient radio emission linked to the optical TDEs) would be very unlikely in a random sample of AGN, suggesting that the radio sources in this sample are unusual for AGN. There is a strong observational correlation between radio AGN activity and SMBH/galaxy mass \citep[e.g.][]{Best2005}, which we do not observe in our sample (although with very low sample statistics). In Figure \ref{fig:sample_corrs} there is no correlation between host galaxy properties, including SMBH mass, and the observed radio luminosities. Overall the compactness and lack of correlation with host properties of our radio-detected sources suggest the radio sources are either very unusual AGN or not AGN at all. This finding suggests the radio sources may be produced by a more exotic mechanism, such as remnant TDE outflow emission or a compact jet. The lack of AGN features in QPE host galaxies is supported at optical wavelengths by a lack of broad emission lines in their optical spectra that is typical of unobscured AGN activity, although in some of the galaxies the narrow line ratios suggest an ionising source in excess of stars \citep[e.g.][]{Wevers2022,Wevers2024}.  


\subsection{Comparison to TDEs in the radio}
The possibility that the radio sources are old synchrotron emitting regions from outflows produced by a TDE is particularly attractive given the apparent link between QPEs and TDEs \citep[e.g.][]{Nicholl2024,Miniutti2025,Bykov2025} and the disc-like quiescent emission observed in many QPE sources which is easily produced by  a disk remaining from a previous TDE \citep[e.g.][]{Linial2023,Franchini2023}. 
Recent studies of optical- and X-ray-selected samples of TDEs have found long-lived radio emission that can be rising up to 10\,yr after the initial TDE \citep{Cendes2024,Goodwin2025}, however, the long-term ($>10$\,yr) radio emission behaviour of TDEs is unknown. 

\begin{figure*}
    \centering
    \includegraphics[width=0.45\columnwidth]{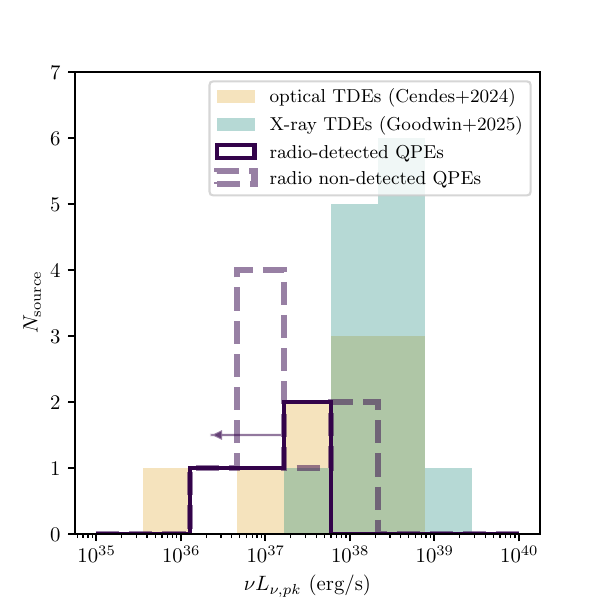}
    \includegraphics[width=0.45\columnwidth]{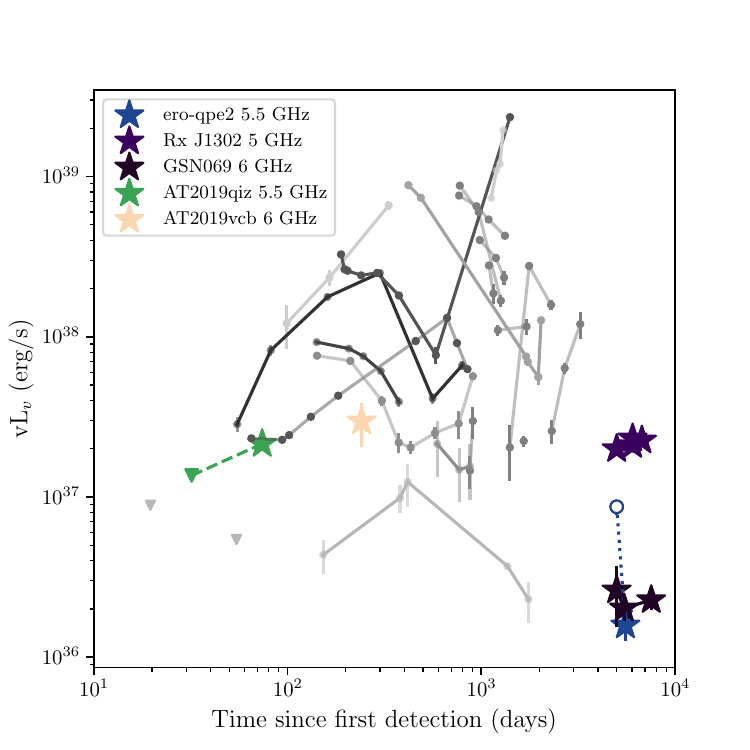}
    \caption{\textit{Left:} The peak observed radio luminosity distribution of the radio-detected QPE sources (solid purple line) and upper limits (dashed purple line). For comparison, the peak observed radio luminosity distribution of the X-ray selected TDE population from \citet{Goodwin2025} is shown in blue and optically-selected TDE population from \citet{Cendes2024} is shown in yellow. The radio luminosity distribution of the QPE sources is broadly consistent with that of TDEs, albeit biased to slightly lower luminosities. \textit{Right:}The observed radio lightcurve of known TDEs (grey) and the five radio-detected QPE sources. Note that eRO-QPE2, GSN\,069, and RxJ1302 are scaled such that the first radio observation is plotted 5000\,d after the first detection since the time the transient emission began is unconstrained. AT2019qiz and AT2019vcb are entirely consistent with the TDE population, while the remaining three QPE sources are less variable than TDEs, although the late-time radio behaviour of TDEs is unknown.}
    \label{fig:tdecomp}
\end{figure*}

In Figure \ref{fig:tdecomp} (left) we plot the observed peak radio luminosity of two TDE samples \citep{Cendes2024,Goodwin2025} and the luminosity distribution of the QPE sources. Broadly the luminosity range of the QPE sources is consistent with TDEs, noting that TDEs span a large range of observed peak luminosities. The radio-detection rate of QPEs (42\%) is consistent with typical radio-detection rates of optical- and X-ray-selected TDEs \citep[40--50\%][]{Cendes2024,Goodwin2025}. 
In Figure \ref{fig:tdecomp} (right) we plot the observed radio lightcurves of the radio-detected QPE sources and those of known radio-emitting TDEs. AT2019qiz and AT2019vcb are consistent with the prompt radio-emitting TDE population. GSN\,069, eRO-QPE2, and RxJ1302 are less variable than the majority of TDEs over time, although it is unclear how variable old radio emission from a synchrotron-emitting outflow would be long after the TDE occurred.  

The radio spectral indices of the QPE sources are consistent with an optically thin synchrotron-emitting source, except AT2019qiz (Section \ref{sec:results}). The radio spectral indices of the QPE sources are also entirely consistent with the radio spectral indices of the X-ray selected TDE sample presented in \citet{Goodwin2025}, which found spectral indices from $\alpha\approx-0.5$ to $\alpha\approx-1.5$ for 11 TDEs. 
In the case of an old, optically thin synchrotron-emitting source with an electron energy index, $p$ of 2.7 \citep[as observed for TDEs, e.g.][]{Cendes2021}, given the \citet{Granot2002} synchrotron spectral model in which the characteristic minimum frequency ($\nu_m$), self-absorption frequency ($\nu_a$) and cooling frequency ($\nu_c$) are ordered $\nu_m < \nu_a < \nu_c$, the spectral index could be either $-p/2=-1.25$ (if the observing frequency is above the cooling break) or $(1-p)/2=-0.85$ (if the observing frequency is below the cooling break). The spectral indices of the radio emission from eRO-QPE2, GSN\,069, and RxJ1302 of $-0.6$ to $-0.9$ are consistent with optically thin synchrotron emission in which the cooling break is above the observed frequency band.
Although, given the age of the QPE sources are as large as $>10$\,yr, it is surprising that the radio spectra do not indicate a steepening due to the cooling break. The synchrotron cooling break evolves with $\nu_c(t)\propto t^{-1/2}$ \citep{Granot2002}. A cooling break was observed in the radio spectrum of the TDE AT2019dsg at 25\,GHz at 83\,d post-TDE \citep{Cendes2021} and 19\,GHz at 132\,d post-TDE for the TDE ASASSN-19bt \citep{Christy2024}. If the cooling break location continues to evolve with time and in the scenario in which the location of the cooling break at time post-TDE is similar for all TDEs, for the cooling break to be above the observed frequency of 5--6\,GHz, the outflow must be $\lesssim250$\,d old. Such a small age can be ruled out for RxJ1302, GSN\,069, and eRO-QPE2 as QPEs have been observed in these systems for longer. However, we note that the radio spectra available for GSN\,069 and eRO-QPE2 are limited in coverage and the synchrotron electron index $p$ may be lower than $p=2.7$ which would result in a less steep spectral index than calculated. Overall, the radio spectra observed for the QPE sources are consistent with radio spectra of TDEs.   

Curiously, \citet{Kosec2025} recently reported the detection of a $1700-2900$\,km/s ionised outflow via X-ray absorption features in observations of GSN\,069. They constrain the outflow distance to be $\approx2-9\times10^{16}$, cm from the SMBH, and do not find any ionized line emission during the QPEs, deducing that the outflow may be linked to recent transient activity in the galaxy nucleus, not to the QPEs themselves. The outflow is similar, although sigificantly faster, to outflows observed in TDEs such as ASASSN-14li \citep{Miller2015} and ASASSN-20qc \citep{Kosec2023}. The outflow in GSN\,069 appears to be stable on very long timescales, unlike the outflow recently detected in ZTF19acknskyy \citep{Chakraborty2025}, which is clearly associated with the QPEs themselves. \citet{Kosec2025} infer that the ionised outflow in GSN\,069 has kinetic power $\dot{E}\sim7\times10-^{39}-2\times10^{41}$\,erg/s from a continuously launched outflow with mass outflow rate $3\times10^{-3}$--$8\times10^{-2}$\,$M_{\odot}$\,yr. They deduce that the outflow cannot be a remnant outflow from previous SMBH activity, being at just 0.03\,pc from the SMBH, and instead conclude the outflow is likely linked to the recent transient activity in GSN\,069 that has been ongoing since 2010. The energy, location, and velocity of this outflow are very similar to constraints obtained on outflows from TDEs in the radio \citep[e.g.][]{Cendes2024}, and synchrotron emission from a shock between this outflow and the circumnuclear medium could produce the radio luminosity of GSN\,069 of $\approx2.5\times10^{36}$\,erg/s. More broadly, this kind of compact ($\sim0.03$\,pc), persistent outflow observed in X-ray for GSN\,069 is consistent with the radio emission seen in all five radio-detected QPE sources, implying that low-level persistent outflows may be common among QPEs and linked to the underlying transient activity that creates the disks observed in these systems, such as a TDE. 


\subsection{Theoretical implications of radio properties for the QPE mechanism}
The lack of strong radio emission in any of the QPE sources and the apparent lack of radio variability associated with the X-ray flares implies that the eruptions themselves do not produce strong outflows or jets. 

\subsubsection{Star-disk collision model}
In the EMRI scenario, current models estimate $\sim10^{-5}$\,M$_{\odot}$ of material is ejected during each disk collision \citep[e.g.][]{Yao2025}, with a velocity $v_{\rm{launch}}$ given by

\begin{equation}
    v_{\rm{launch}} = c \sqrt{\frac{R_g}{R_{\rm{collision}}}}
\end{equation}
where $R_g$ is the gravitational radius of the SMBH and $R_{\rm{collision}}$ the radius at which the object collides with the disk. \citet{Chakraborty2025} detected rapidly varying absorption features in ZTF19acknskyy associated with blueshifted emission corresponding to $v_{\rm{out}}\sim6-40\%\,c$, consistent with $R_{\rm{collision}}$ of 2.5--15 $R_g$ for a $10^{6}$\,M$_{\odot}$ SMBH.
The outflow would then propagate freely until it encounters circumnuclear gas which may produce a shock that would emit synchrotron radiation. If we assume the radius at which this shock occurs is $R_{\rm{shock}}$, then the velocity of the outflow at which time the shock occurs is
\begin{equation}
    v_{\rm{shock}} = c \sqrt{\frac{R_g}{R_{\rm{shock}}}} 
\end{equation}

For a $10^{6}$\,M$_{\odot}$ SMBH and a shock radius of $\sim10^{16}$\,cm \citep[typical of early TDE outflow emission e.g.][]{Alexander2016,Goodwin2022} we find $v_{\rm{shock}}\sim0.005\,c$. 
Assuming there is $m_{\rm{ej}}\sim10^{-5}$\,M$_{\odot}$ mass in the outflow with velocity 0.005$\,c$, such an outflow would have energy $E = \frac{1}{2}m_{\rm{ej}}v_{\rm{shock}}^2$ $\sim10^{44}$\,erg. This energy is many orders of magnitude smaller than typical outflow energies derived for TDEs of $10^{49}-10^{51}$\,erg \citep[e.g.][]{Goodwin2025,Cendes2024}, and would not produce synchrotron emission with sufficient radio luminosity to detect, even in the case of an extremely dense circumnuclear medium or strong magnetic field. Therefore, the lack of variable radio emission detected correlated with the X-ray QPEs is consistent with the outflows expected in the disk collision model. 

\subsubsection{Disk instability model}

In the disk instability model for QPEs, the X-ray flares are suggested to be produced by instability cycles where the QPEs are caused by magnetically driven outflows \citep[e.g.][]{Pan2022,Pan2023}. It is unclear in this scenario if episodic jet activity from each accretion event would be expected or detectable, as this would depend on the accretion rate, efficiency, and mass accreted during each instability event. However, the radio properties of QPEs observed to date would suggest that such instabilities do not result in significant outflows or jets, which one may expect if the instabilities trigger short and low-mass accretion events onto the SMBH.

\section{Summary}\label{sec:summary}

In this work we present a comprehensive analysis of the radio properties of the 12 known QPE sources. Overall we find weak, compact radio emission coincident with 5/12 of the QPE sources. We find no correlation between the radio luminosity and QPE flare properties such as recurrence time, duration, and total energy radiated or the host galaxy properties such as SMBH mass, galaxy mass, or star formation rate. 
The radio-detected QPE sources are compact $<2.3$\,kpc, localised to the center of their host galaxies, and faint ($\nu L_{\nu}\sim5\times10^{36}-5\times10^{38}$\,erg/s). The radio spectra are optically thin (except AT2019qiz which shows a young radio-emitting outflow likely produced by the optically-detected TDE), with spectral indices consistent with AGN or TDE populations. The compactness of the radio sources and lack of correlation between radio luminosity and SMBH mass is very unusual for AGN, suggesting the radio sources may instead to be linked to more exotic transient activity in the nuclei of the galaxies, such as a relatively recent TDE. The radio spectra and luminosities of the QPE sources are broadly consistent with observed radio emission from TDEs, although the lack of a long-term radio decay in GSN\,069 is unexpected for an ageing TDE outflow. Future radio observations of a larger sample of QPEs will determine if compact, weak radio sources are common among the population of QPE sources, and may provide further insight into the mechanism driving the radio emission observed in the current sample.   

\begin{acknowledgement}
AJG is grateful for support from the Forrest Research Foundation. GM acknowledges MICIU/AEI/10.13039/501100011033 for support through grants n. PID2020-115325GB-C31 and PID2023-147338NB-C21. R.A. was supported by NASA through the NASA Hubble Fellowship grant \#HST-HF2-51499.001-A awarded by the Space Telescope Science Institute, which is operated by the Association of Universities for Research in Astronomy, Incorporated, under NASA contract NAS5-26555.
 
The Australia Telescope Compact Array is part of the Australia Telescope National Facility (grid.421683.a) which is funded by the Australian Government for operation as a National Facility managed by CSIRO. We acknowledge the Gomeroi people as the traditional owners of the Observatory site. The National Radio Astronomy Observatory is a facility of the National Science Foundation operated under cooperative agreement by Associated Universities, Inc.
\end{acknowledgement}



\appendix

\section{ATCA radio observations of eRO-QPE2}\label{sec:eRO-QPE2}

\begin{figure}
    \centering
    \includegraphics[width=\columnwidth]{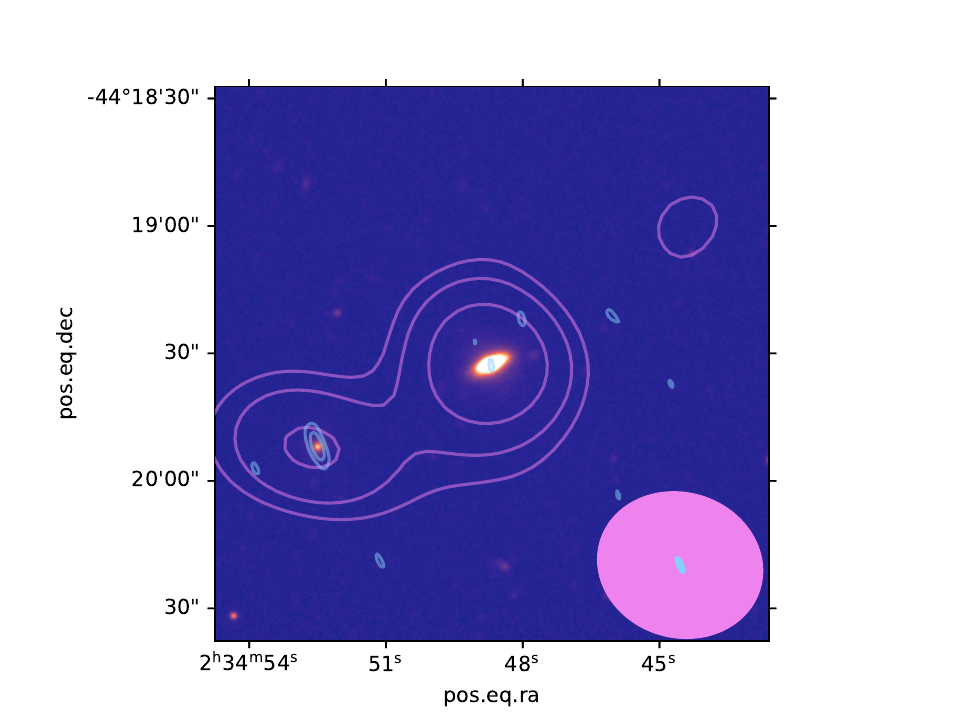}
    \caption{The DESI Legacy DR8 optical image of 2MASX J02344872-4419325, the host galaxy of eRO-QPE2, and the ATCA 5.5\,GHz radio contours overlaid. The pink contours show the lower resolution ATCA observations from June 2022 whereas the blue contours show the higher resolution ATCA observations from December 2022. The size of the beam for each radio observation is plotted in the bottom right corner indicating the nominal image resolution.}
    \label{fig:eroqpe2_radcontours}
\end{figure}

The first ATCA radio observation of eRO-QPE2 was taken with the ATCA in compact H214 configuration, resulting in a low image resolution. There are two nearby sources which eRO-QPE2 is confused with in the compact configuration observation, resulting in a higher flux density measured for the target radio source. The second ATCA observation was taken in the extended 6\,km configuration with a much higher image resolution. In Figure \ref{fig:eroqpe2_radcontours} we plot the optical image of the field with radio contours from both of the ATCA observations. It is clear that only in the second ATCA observation are the sources not confused. We therefore caution readers from interpreting the change in flux density between the two ATCA epochs as intrinsic source variability. 

\section{Radio Observations}

All publicly available and new radio observations presented in this work of QPE sources are given in Table \ref{tab:radio_obs}. 

\clearpage
\onecolumn
\begin{longtable}{p{2.5cm}p{2cm}p{2cm}p{2.5cm}p{2cm}p{2.5cm}p{5cm}}
    \label{tab:radio_obs}\\
    \caption{Radio observations of QPE sources.}\\
    \hline
    Source & Date & Instrument & Program ID & Frequency (GHz) & Flux Density (${\mu}$Jy) & Ref. \\
\hline
AT2019qiz & 21/10/2019 & ATCA & CX442 & 5.5 & $<$450 & ATel\#13310\\
AT2019qiz & 21/10/2019 & ATCA & CX442 & 9.0 & 210$\pm$80 & ATel\#13310\\
AT2019qiz & 17/11/2019 & ATCA & CX442 & 17.0 & 1350$\pm$400 & ATel\#13310\\
AT2019qiz & 17/11/2019 & ATCA & CX442 & 19.0 & 1240$\pm$400 & ATel\#13310\\
AT2019qiz & 2/12/2019 & ATCA & CX442 & 5.5 & 710$\pm$40 & ATel\#13334\\
AT2019qiz & 2/12/2019 & ATCA & CX442 & 9.0 & 1690$\pm$30 & ATel\#13334\\
AT2019qiz & 2/12/2019 & ATCA & CX442 & 16.7 & 2380$\pm$40 & ATel\#13334\\
AT2019qiz & 2/12/2019 & ATCA & CX442 & 21.2 & 2610$\pm$50 & ATel\#13334\\
\hline
eRO-QPE1 & 11/9/2021 & MeerKAT & DDT-20210908- RA-01 & 1.28 & $<$27 & Chakraborty+2024\\
eRO-QPE1 & 19/9/2021 & MeerKAT & DDT-20210908- RA-01 & 1.28 & $<$30 & Chakraborty+2024\\
eRO-QPE1 & 19/1/2023 & VLA & 23A–059 & 6.0 & $<$32 & Chakraborty+2024\\
eRO-QPE1 & 20/1/2023 & VLA & 23A–059 & 6.0 & $<$38 & Chakraborty+2024\\
eRO-QPE1 & 23/1/2023 & VLA & 23A–059 & 6.0 & $<$36 & Chakraborty+2024\\
eRO-QPE1 & 24/1/2023 & VLA & 23A–059 & 6.0 & $<$88 & Chakraborty+2024\\
eRO-QPE1 & 27/1/2023 & VLA & 23A–059 & 6.0 & $<$41 & Chakraborty+2024\\
eRO-QPE1 & 29/1/2023 & VLA & 23A–059 & 6.0 & $<$33 & Chakraborty+2024\\
eRO-QPE1 & 30/1/2023 & VLA & 23A–059 & 6.0 & $<$34 & Chakraborty+2024\\
\hline
eRO-QPE2 & 21/6/2022 & ATCA-H214 & CX507 & 5.0 & $<$214$^*$ & this work\\
eRO-QPE2 & 21/6/2022 & ATCA-H214 & CX507 & 6.0 & $<$168$^*$ & this work\\
eRO-QPE2 & 21/6/2022 & ATCA-H214 & CX507 & 8.5 & $<$211$^*$ & this work\\
eRO-QPE2 & 21/6/2022 & ATCA-H214 & CX507 & 9.5 & $<$114$^*$ & this work\\
eRO-QPE2 & 21/12/2023 & ATCA-6 & C3549 & 2.1 & $<$130 & this work\\
eRO-QPE2 & 18/12/2023 & ATCA-6 & C3549 & 5.5 & 39$\pm$8 & this work\\
eRO-QPE2 & 18/12/2023 & ATCA-6 & C3549 & 9.0 & 25$\pm$9 & this work\\
eRO-QPE2 & 10/1/2024 & ATCA-6 & C3549 & 5.5 & $<$44 & this work\\
eRO-QPE2 & 10/1/2024 & ATCA-6 & C3549 & 9.0 & $<$37 & this work\\
\hline
eRO-QPE3 & 1/10/2022 & ATCA & C3513 & 5.5 & $<$54 & Arcodia+2024\\
eRO-QPE3 & 1/10/2022 & ATCA & C3513 & 9.0 & $<$49 & Arcodia+2024\\
eRO-QPE3 & 13/6/2023 & ATCA & C3513 & 5.5 & $<$51 & Arcodia+2024\\
eRO-QPE3 & 13/6/2023 & ATCA & C3513 & 9.0 & $<$49 & Arcodia+2024\\
\hline
eRO-QPE4 & 5/5/2023 & ATCA-1.5A & C3527 & 5.5 & $<$48 & Arcodia+2024\\
eRO-QPE4 & 5/5/2023 & ATCA-1.5A & C3527 & 9.0 & $<$39 & Arcodia+2024\\
eRO-QPE4 & 20/8/2023 & ATCA-6D & C3527 & 2.1 & $<$298 & Arcodia+2024\\
\hline
eRO-QPE5 & 26/4/2024 & ATCA-6A & C3586 & 5.5 & $<$36 & Arcodia+2025\\
eRO-QPE5 & 26/4/2024 & ATCA-6A & C3586 & 9.0 & $<$33 & Arcodia+2025\\
\hline
GSN069 & 2/11/2017 & VLA & 17B-027 & 6.0 & 61$\pm$25 & Miniutti+2019\\
GSN069 & 14/2/2019 & VLA & 19A-454 & 6.0 & 47$\pm$8 & Miniutti+2019\\
GSN069 & 26/1/2019 & ATCA & CX425 & 5.5 & $<$126 & Miniutti+2019\\
GSN069 & 26/1/2019 & ATCA & CX425 & 9.0 & $<$129 & Miniutti+2019\\
GSN069 & 15/2/2019 & ATCA & CX425 & 5.5 & $<$126 & Miniutti+2019\\
GSN069 & 15/2/2019 & ATCA & CX425 & 9.0 & $<$120 & Miniutti+2019\\
GSN069 & 14/2/2019 & MeerKAT & 1550159669 & 1.3 & 147$\pm$7 & Miniutti+2019\\
GSN069 & 15/2/2019 & MeerKAT & 1550159669 & 1.3 & 156$\pm$14 & Miniutti+2019\\
GSN069 & 16/10/2024 & ATCA & C3622 & 2.1 & 162$\pm$15 & this work\\
GSN069 & 16/10/2024 & ATCA & C3622 & 5.5 & 53$\pm$7 & this work\\
GSN069 & 16/10/2024 & ATCA & C3622 & 9.0 & 41$\pm$7 & this work\\
\hline
RX J1301.9+2747 & 5/7/2015 & VLA & 15A-349 & 9.0 & 119$\pm$10 & Yang+2022\\
RX J1301.9+2747 & 7/8/2015 & VLA & 15A-349 & 9.0 & 155$\pm$10 & Yang+2022\\
RX J1301.9+2747 & 7/8/2015 & VLA & 15A-349 & 9.0 & 123$\pm$10 & Yang+2022\\
RX J1301.9+2747 & 7/8/2015 & VLA & 15A-349 & 9.0 & 142$\pm$10 & Yang+2022\\
RX J1301.9+2747 & 8/8/2015 & VLA & 15A-349 & 9.0 & 78$\pm$11 & Yang+2022\\
RX J1301.9+2747 & 9/8/2015 & VLA & 15A-349 & 9.0 & 89$\pm$11 & Yang+2022\\
RX J1301.9+2747 & 10/8/2015 & VLA & 15A-349 & 9.0 & 90$\pm$10 & Yang+2022\\
RX J1301.9+2747 & 10/8/2015 & VLA & 15A-349 & 9.0 & 145$\pm$10 & Yang+2022\\
RX J1301.9+2747 & 4/9/2017 & VLA & 17B-027 & 6.0 & 295$\pm$10 & Yang+2022\\
RX J1301.9+2747 & 5/1/2019 & VLA & 18B-115 & 14.0 & 184$\pm$9 & Yang+2022\\
RX J1301.9+2747 & 6/1/2019 & VLA & 18B-115 & 14.0 & 163$\pm$12 & Yang+2022\\
RX J1301.9+2747 & 7/1/2019 & VLA & 18B-115 & 14.0 & 189$\pm$9 & Yang+2022\\
RX J1301.9+2747 & 8/1/2019 & VLA & 18B-115 & 14.0 & 177$\pm$9 & Yang+2022\\
RX J1301.9+2747 & 9/1/2019 & VLA & 18B-115 & 14.0 & 183$\pm$8 & Yang+2022\\
RX J1301.9+2747 & 10/1/2019 & VLA & 18B-115 & 14.0 & 148$\pm$9 & Yang+2022\\
RX J1301.9+2747 & 12/1/2019 & VLA & 18B-115 & 14.0 & 179$\pm$9 & Yang+2022\\
RX J1301.9+2747 & 31/5/2015 & GMRT & 28\_039 & 1.4 & 764$\pm$23 & Yang+2022\\
RX J1301.9+2747 & 14/2/2017 & VLBA & BS255 & 1.6 & 670$\pm$33 & Yang+2022\\
RX J1301.9+2747 & 18/6/2006 & VLA & AM868 & 1.4 & 866$\pm$117 & Yang+2022\\
RX J1301.9+2747 & 17/10/2020 & ASKAP & RACs & 0.89 & 1880$\pm$348 & this work\\
RX J1301.9+2747 & 11/7/2020 & VLA & SJ6456 & 6.0 & 313$\pm$2 & Giustini+2024\\
RX J1301.9+2747 & 12/7/2020 & VLA & SJ6456 & 6.0 & 344$\pm$3 & Giustini+2024\\
RX J1301.9+2747 & 17/6/2022 & VLA & SL0464 & 6.0 & 334$\pm$3 & Giustini+2024\\
\hline
ZTF19acnskyy & 27/7/2024 & ATCA & CX572 & 5.5 & $<$84 & Hernandez-Garcia+2025\\
ZTF19acnskyy & 27/7/2024 & ATCA & CX572 & 9.0 & $<$93 & Hernandez-Garcia+2025\\
\hline
AT2022upj & 26/3/2023 & VLA & VLASS & 3.0 & $<$472 & this work\\
AT2022upj & 12/11/2017 & VLA & VLASS & 3.0 & $<$410 & this work\\
\hline
XMMSLJ024916.6-041244 & 12/5/2023 & VLA & VLASS & 3.0 & $<$437 & this work\\
XMMSLJ024916.6-041244 & 13/9/2020 & VLA & VLASS & 3.0 & $<$465 & this work\\
XMMSLJ024916.6-041244 & 28/11/2017 & VLA & VLASS & 3.0 & $<$403 & this work\\
         \hline
AT2019vcb & 30/4/2020 & VLA & 20A-392 & 10.0 & 49$\pm$11 & this work\\
AT2019vcb & 12/5/2020 & VLA & 20A-392 & 10.0 & 30$\pm$9 & this work\\
AT2019vcb & 13/7/2020 & VLA & 20A-392 & 15.0 & $<$25 & this work\\
AT2019vcb & 13/7/2020 & VLA & 20A-392 & 10.0 & $<$23 & this work\\
AT2019vcb & 13/7/2020 & VLA & 20A-392 & 6.0 & 24$\pm$8 & this work\\
AT2019vcb & 13/7/2020 & VLA & 20A-392 & 3.0 & 42$\pm$10 & this work\\
AT2019vcb & 19/7/2020 & VLA & 20A-392 & 15.0 & $<$19 & this work\\
AT2019vcb & 19/7/2020 & VLA & 20A-392 & 10.0 & $<$19 & this work\\
AT2019vcb & 19/7/2020 & VLA & 20A-392 & 6.0 & 23$\pm$8 & this work\\
AT2019vcb & 19/7/2020 & VLA & 20A-392 & 3.0 & 45$\pm$9 & this work\\
         \hline
         \hline
\end{longtable}
\footnotesize{$^*$ We treat this flux density measurement as an upper limit due to the target being confused with a nearby source.}

\twocolumn
\bibliography{bibfile}


\end{document}